\documentclass{PoS}
\usepackage{natbib,natbibspacing}
\usepackage{graphicx}
\usepackage{epsfig}
\usepackage{amssymb}
\newcommand{\dnin}{nW m$^{-2}$ sr$^{-1}$}

\title{Measuring Light from the Epoch of Reionization with CIBER, the Cosmic Infrared Background Experiment}

\ShortTitle{Measuring Light from the Epoch of Reionization with CIBER, the Cosmic Infrared Background Experiment}

\author{\speaker{Michael Zemcov} \\
       California Institute of Technology/Jet Propulsion Laboratory\\
       E-mail: \email{zemcov@caltech.edu}}

\author{T.~Arai$^{\mathrm{a}}$,
  J.~Battle$^{\mathrm{b}}$, 
  J.~Bock$^{\mathrm{b,c}}$, 
  A.~Cooray$^{\mathrm{d}}$, 
  V.~Hristov$^{\mathrm{c}}$, 
  B.~Keating$^{\mathrm{e}}$,
  M.~G.~Kim$^{\mathrm{f}}$,
  D.~H.~Lee$^{\mathrm{g}}$, 
  L.~Levenson$^{\mathrm{c}}$, 
  P.~Mason$^{\mathrm{c}}$, 
  T.~Matsumoto$^{\mathrm{f}}$, 
  S.~Matsuura$^{\mathrm{h}}$,
  K.~Mitchell-Wynne$^{\mathrm{d}}$,
  U.~W.~Nam$^{\mathrm{g}}$, 
  T.~Renbarger$^{\mathrm{e}}$, 
  J.~Smidt$^{\mathrm{d}}$,
  I.~Sullivan$^{\mathrm{i}}$, 
  K.~Tsumura$^{\mathrm{h}}$, and
  T.~Wada$^{\mathrm{h}}$\\
$^{\mathrm{a}}$University of Tokyo;
$^{\mathrm{b}}$Jet Propulsion Laboratory;
$^{\mathrm{c}}$California Institute of Technology;
$^{\mathrm{d}}$University of California, Irvine;
$^{\mathrm{e}}$University of California, San Diego;
$^{\mathrm{f}}$Seoul National University;
$^{\mathrm{g}}$Korea Astronomy and Space Science Institute;
$^{\mathrm{h}}$Japan Aerospace Exploration Agency;
$^{\mathrm{i}}$University of Washington\\
}

\abstract{Ultraviolet emission from the first generation of stars in
  the Universe ionized the intergalactic medium in a process which was
  completed by $z \sim 6$; the wavelength of these photons has been
  redshifted by $(1+z)$ into the near infrared today and can be measured
  using instruments situated above the Earth's atmosphere.  First
  flying in February 2009, the Cosmic Infrared Background Experiment
  (CIBER) comprises four instruments housed in a single reusable
  sounding rocket borne payload.  CIBER will measure spatial
  anisotropies in the extragalactic IR background caused by
  cosmological structure from the epoch of reionization using two
  broadband imaging instruments, make a detailed characterization of
  the spectral shape of the IR background using a low resolution
  spectrometer, and measure the absolute brightness of the Zodical
  light foreground with a high resolution spectrometer in each of our
  six science fields.  This paper presents the scientific
  motivation for CIBER and details of its first two flights, including
  a review of the published scientific results from the first flight and an
  outlook for future reionization science with CIBER data.}

\FullConference{Cosmic Radiation Fields: Sources in the early Universe\\
		 November 9-12, 2010\\
		 DESY, Germany}

\begin{document}

\section{Introduction}
\label{S:intro}

The extragalactic background light (EBL) is the sum of all of the
light emitted throughout the history of the Universe.  At near
infrared (IR) wavelengths, the EBL is predominantly due to stellar
emission from nucleosynthesis (see \citealt{Hauser2001} for a review).
However, absolute measurements of the EBL at near IR wavelengths to
date lack the systematic error control to constrain the radiative
content of the cosmos to within an order of magnitude
(\citealt{Hauser1998}, \citealt{Dwek1998}, \citealt{Gorjian2000},
\citealt{Wright2001}, \citealt{Cambresy2001}, \citealt{Matsumoto2005},
\citealt{Levenson2007}).  Further, the summed contribution of galaxies to
the EBL does not reproduce the EBL measured by absolute photometric
instruments.  For example, at $\lambda = 3.6 \, \mu$m the EBL measured
by DIRBE from absolute photometry is $12.4 \pm 3.2 \,$nW m$^{-2}$
sr$^{-1}$ \citep{Wright2000}, while the deepest pencil beam surveys
with Spitzer give $6{-}9 \,$nW m$^{-2}$ sr$^{-1}$ (\citealt{Fazio2004},
\citealt{Sullivan2007}).  At shorter wavelengths this divergence is even
more pronounced.

The discrepancy between absolute photometric measurements and
integrated number counts leaves open the possibility that there exists
some truly diffuse emission comprising a fraction of the near IR
background and that using suitable instruments it may be possible to
measure it.  Importantly, as the EBL traces star formation throughout
the history of the Universe, it contains information about the
earliest generation of stars which were responsible for ionizing it
(see \citealt{Fan2006} for a review).  Though the spectrum of the near IR
EBL does contain information about the epoch of reionization, it is
expected to be faint compared to the contribution from galaxies.
However, as the signature from the first stars is expected to be both
diffuse and structured on large scales, the spatial fluctuations of
its imprint on the near IR EBL yield a great deal of information about
the formation of structure during the time of the first stars
\citep{Cooray2004}.

The Cosmic Infrared Background Experiment (CIBER) is a series of
sounding-rocket borne instrument payloads designed to measure both the
spectrum of the EBL and the spatial fluctuations in the EBL imprinted
during the epoch of reionization (REBL).  The first CIBER payload
configurations will probe the power spectrum of near IR EBL
fluctuations for an REBL component, limit the strength of the Lyman
cutoff signature of reionization in the near IR EBL between the
optical and near-infrared EBL measurements, and measure the EBL from
$0.7{-}2.1 \, \mu$m down to the zodiacal foreground subtraction limit.
This paper reviews the scientific motivation for CIBER, gives a brief
discussion of the instrumentation package, and finally discusses the
outlook for measurement of the epoch of reionization using the near IR
EBL over the coming decade.

\section{Scientific Motivation for CIBER}
\label{S:science}

The epoch of reionization critically shapes our understanding of the
formation of structure in the Universe.  During this time and the
cosmological dark ages which preceded it, matter collapsed under
gravity into over-dense regions seeded in the early Universe to form
the first stars and galaxies.  Unfortunately, these epochs are also
the most difficult to study as the electromagnetic radiation produced
by the matter during this process was efficiently reabsorbed by the
ubiquitous neutral medium.  The signals from this epoch are faint and
that the matter in the Universe which has been radiating since that
reionization generates a strong foreground to the measurement.

As a result, the epoch of reionization largely remains an
observational enigma.  Fortunately, the physical processes involved in
reionization are well understood: since the early Universe contained
only an electrically neutral gas composed predominantly of hydrogen,
it was opaque to rest frame photons shorter than the Lyman-$\alpha$
transition at $122 \,$nm.  Such ultraviolet (UV) photons were
efficiently absorbed by electrons occupying the ground state of
hydrogen so that such photons had a very small mean free path, which
is to say, photons were absorbed so quickly that neighbouring regions
were invisible to one another.  The cosmological dark ages ended when
the first generation of stars and their associated stellar products
and remnants ionized the neutral medium in the early Universe in a
relatively brief epoch which present measurements show was underway by
a redshift $z \gtrsim 10$ and almost complete by $z \sim 5$
(\citealt{Komatsu2010}, \citealt{Bouwens2010}).

Since cosmological redshifting increases the wavelength of photons,
the UV photons emitted by these earliest stars are now observable in
the near IR.  This has been used by very deep pencil-beam surveys to
find the highest known redshift galaxies (see
e.g.~\citealt{Bouwens2010} and references therein) which are thought
to represent the population of (proto-)galaxies responsible for
reionization.  Importantly, when the energy required to reionize the
Universe is calculated, it is found that many more photons are
necessary than can be produced by the observed sources alone.  This is
perhaps not surprising: the detected very high redshift sources belong
to the rare, bright population of the source number counts and, as at
later epochs, the exponentially more fainter sources present at
reionization dominate the integrated light.  The implication is that
the Universe was reionized by many faint sources rather than few
bright ones.

Current work has gone some of the way towards detecting these sources,
but clearly much more of the integrated light has yet to be resolved
into discrete sources and these observations are pushing the limits of
the current generation of instruments.  In fact, the results of
\citet{Salvaterra2010} suggest that even next generation telescopes
like the James Webb Space Telescope (JWST) will have difficultly
resolving more than half of the sources responsible for the integrated
light for $z>6$.  Clearly, other avenues which allow measurement of
the light from reionization in the near IR should be pursued.

\begin{figure}
\centering
\epsfig{file=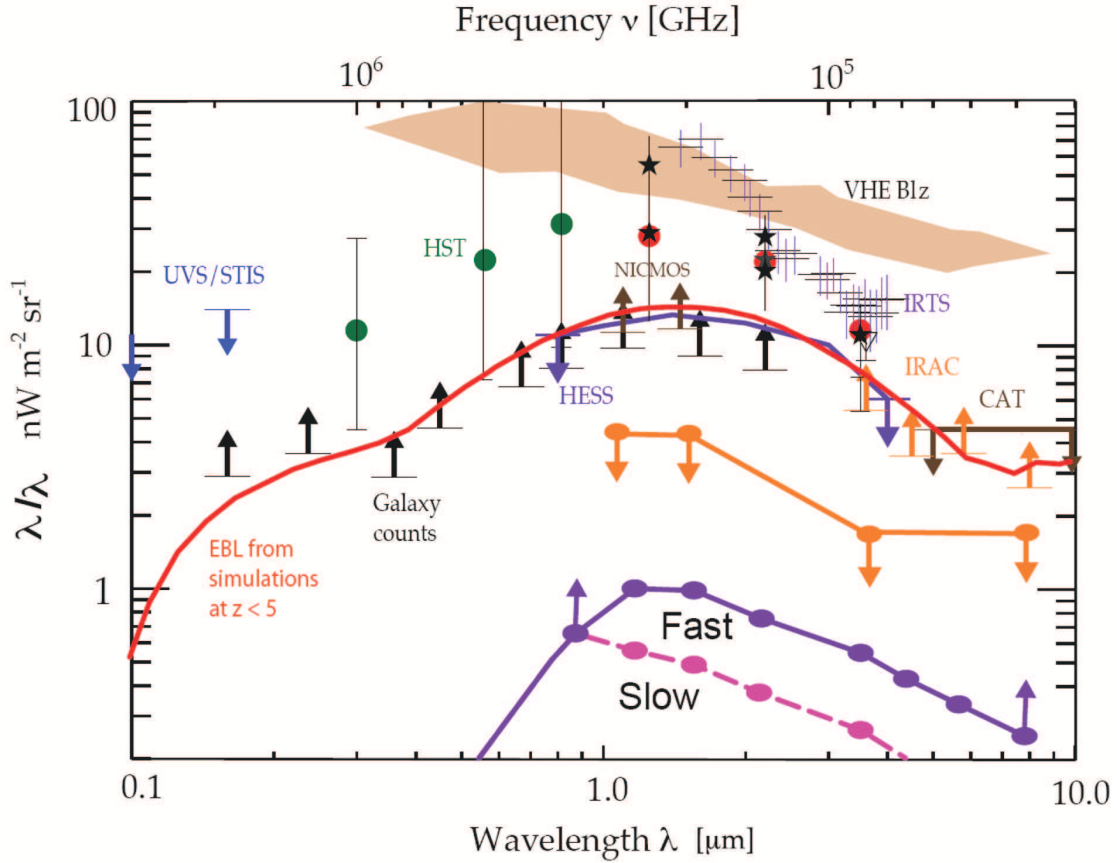,width=0.99\textwidth}
\caption{Measurements and expectations for the spectrum of the
  optical/IR EBL including estimates of the contribution from
  reionization.  The points above $\sim 10 \,$\dnin\ show current
  estimates of the total EBL from instruments capable of absolute
  photometry, while the points shown as lower limits show the EBL
  inferred from integrations of galaxy counts.  Neither currently
  strongly constrain some contribution from diffuse emission such as
  would be produced by reionization.  The pink and purple lines
  labelled ``fast'' and ``slow'' show the expected spectral energy
  distribution from two models of reionization; these are conservative
  lower limits which allow for only a minimum REBL.  The yellow upper
  limits are current constraints on the level of the reionization
  signal from fluctuations in the near IR EBL light.}
\label{fig:ebl}
\end{figure}

It has been known for some time that galaxies at the epoch of
reionization will contribute to the EBL (\citealt{Santos2002},
\citealt{Salvaterra2003}), though the estimates of the fractional
contribution to the overall EBL vary.  As this light was produced by
faint, nebulous sources at a shell centered at the observer, this
would look like a diffuse component to the EBL and not readily resolve
into individual sources.  In fact, current measurements of the EBL do
not preclude the presence of such a diffuse signal and allow
considerable room between the integrated light from number counts and
absolute photometric measurements; the current state of the art is
shown in Figure \ref{fig:ebl}.  Given tight enough controls on the
systematic errors in the measurement, current absolute photometric
instruments have more than sufficient sensitivity to measure the first
light component of the EBL.

Unfortunately, the systematic errors affecting such a measurement are
formidable and so far have proved to be the limiting factor in such a
search.  This is primarily due to the astrophysical foregrounds to the
measurement; not only do observers in the vicinity of earth have to
contend with the Zodiacal light\footnote{The dominant large scale
  emission in the near IR band near the Earth is the Zodiacal light
  (ZL); this diffuse sky brightness is due to sunlight scattering off
  small dust particles lying in the plane of the ecliptic.  Though it
  is only structured on large scales and does not directly affect
  measurements of fluctuation power from REBL, it is a major concern
  for absolute photometric measurements of the overall EBL.}, which is
$> 1000$ times brighter than the REBL at any wavelength,
but also the EBL emitted by all galaxies \emph{after} reionization is a
foreground to the measurement.  Disentangling these is a difficult
prospect and requires careful design of both the instrument and
observation strategy to optimize the chance of success.

\begin{figure}
\centering
\epsfig{file=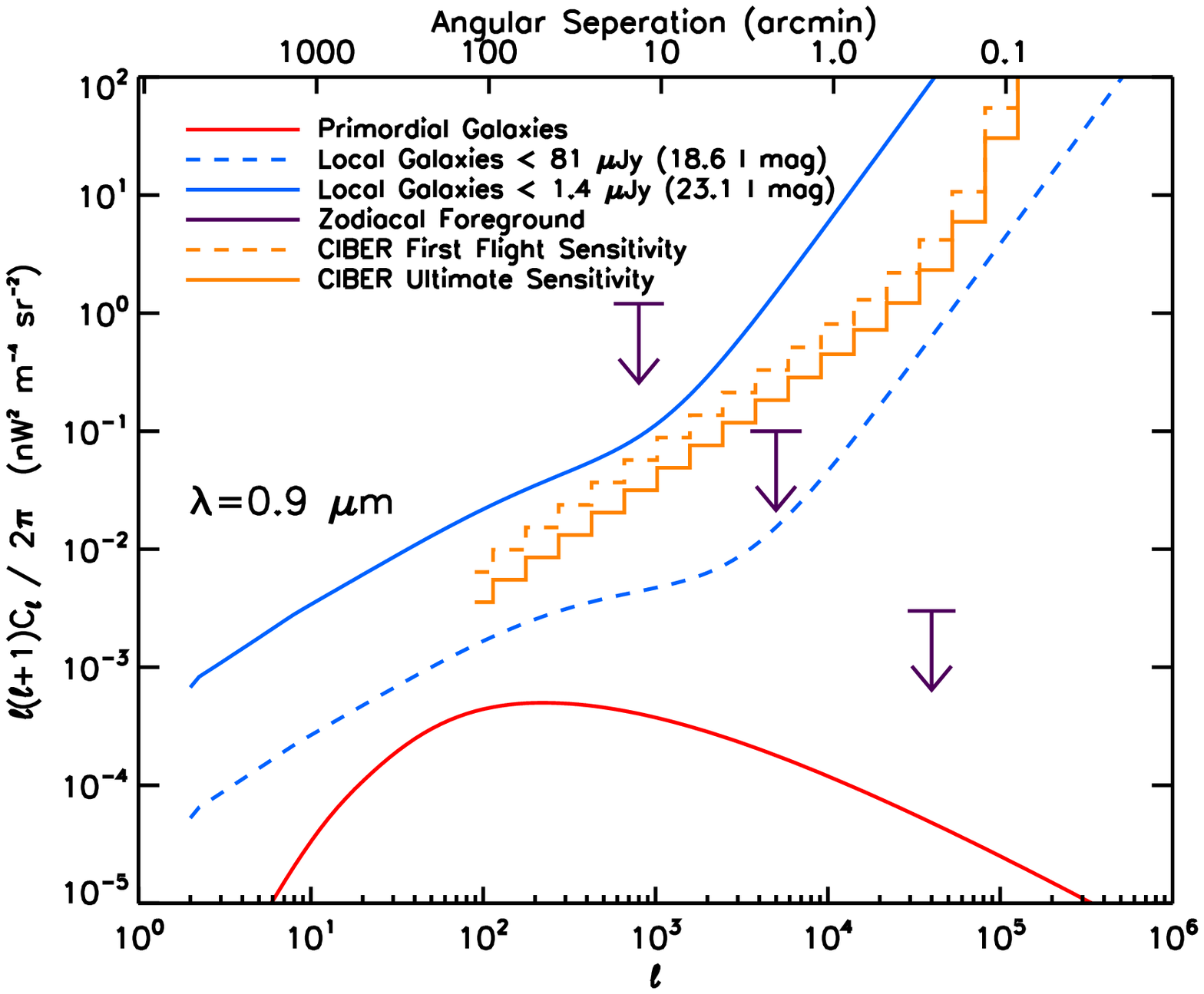,width=0.49\textwidth}
\epsfig{file=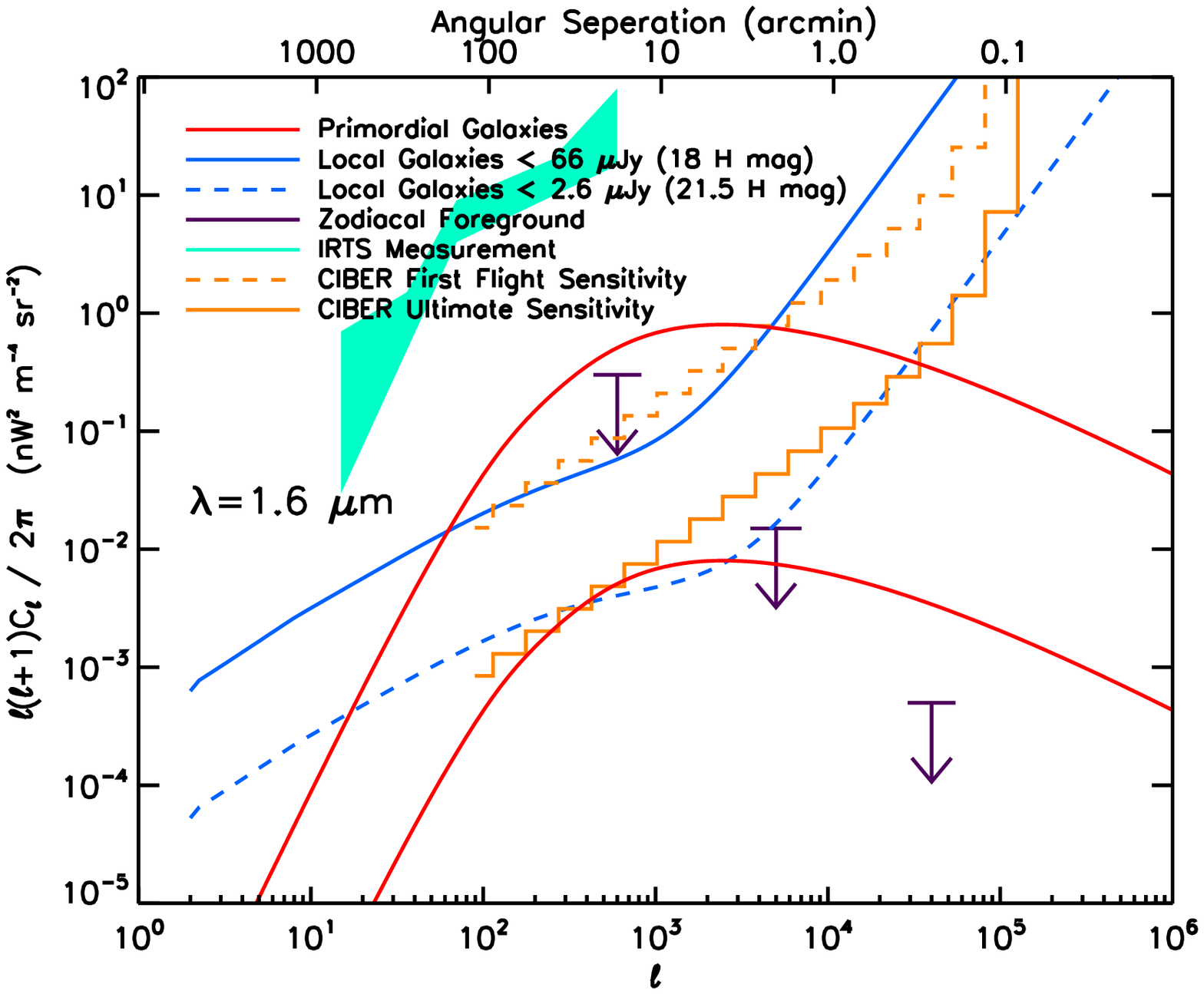,width=0.49\textwidth}
\epsfig{file=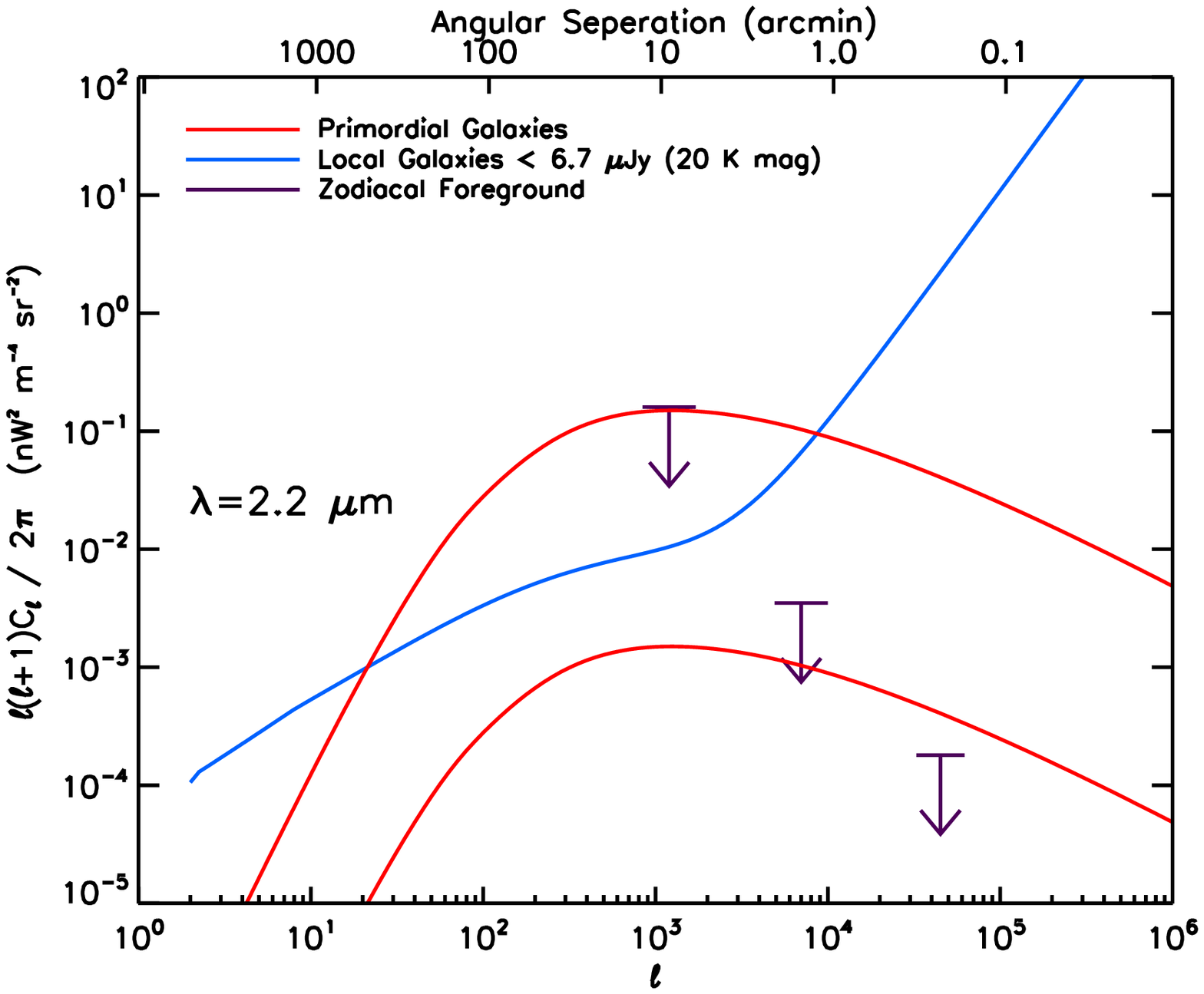,width=0.49\textwidth}
\epsfig{file=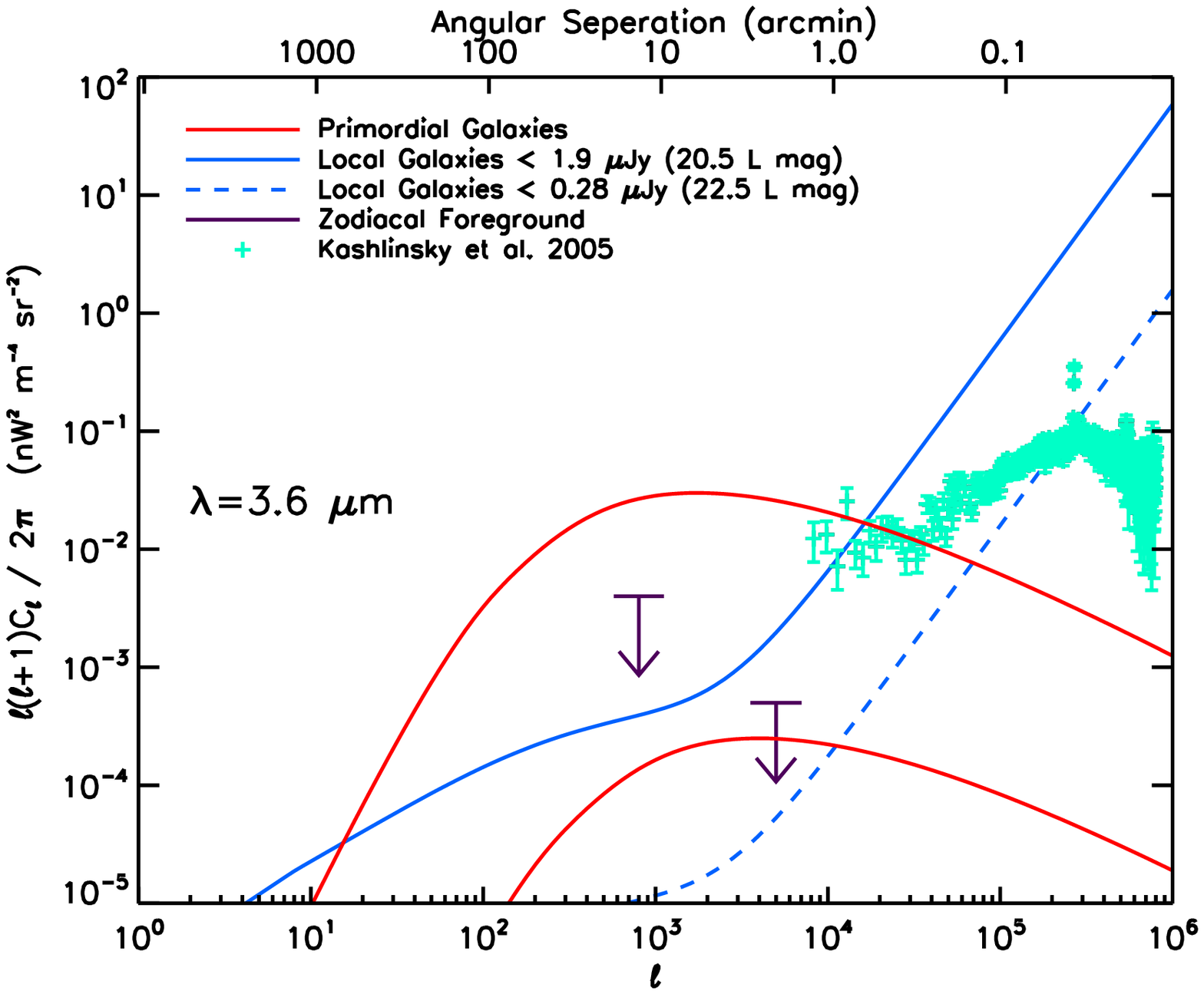,width=0.49\textwidth}
\caption{Spatial power spectrum of EBL fluctuations in standard IR
  bands. The red curves show the power spectra of fluctuations from
  first-light galaxies forming over the redshift interval 8 < z < 15.
  The bottom solid line shows the (approximate) minimal signal
  necessary for reionization to proceed.  Fluctuation measurements are
  shown for Spitzer-IRAC deep, narrow fields as cyan points
  \citep{Kashlinsky2005}.  The blue curves give estimated
  fluctuations from known galaxies, as a function of magnitude cutoff,
  based on a galaxy distribution model matched to existing clustering
  data.  The galaxy fluctuation signal is a combination of clustering,
  dominant on large scales, and shot noise, increasing as $\ell^{2}$
  and dominant on small scales.  The galaxy cutoff taken for CIBER is
  25\% pixel removal using deep ancillary source catalogs.  The orange
  stairsteps show the $\delta \ell / \ell$-binned statistical
  sensitivity of CIBER (in 2 bands) in a single 50s observation, while
  the dashed line shows the statistical sensitivity achieved in the
  first flight in February 2009.  Even this degraded level of
  performance should be adequate to test if the fluctuations seen by
  Spitzer-IRAC are due to first-light galaxies at z > 8.  Zodiacal
  light is known to be spatially uniform, shown by the upper limits
  based on Spitzer and Akari scaled from mid-IR wavelengths. }
\label{fig:irbps}
\end{figure}

A measurement from which a great deal can be learned about
reionization and requires somewhat less stringent control of
systematic errors is to measure fluctuations in the near IR
background.  These fluctuations are due to variations in the density
of reionizing sources from place to place in the Universe and can be
modelled using our understanding of the matter power spectrum over
time and the physical processes involved.  Indeed, the power spectra
of such fluctuations have been calculated by \citet{Cooray2004} and
\citet{Fernandez2010} using different methods which give broadly
compatible results.  The power spectra of \citet{Cooray2004} are shown
in Figure \ref{fig:irbps}\ at four different near IR astronomical
bands.  Definitive detection of such power spectra would allow two key
features of reionization to be studied.  First, the amplitude of the
power spectral signature probes the integrated star formation rate
during reionization.  Second, the width of this reionization
signature's spectrum probes the redshift duration of reionization.
These measurements are complimentary to information from CMB
polarization and 21 cm background studies of neutral hydrogen and
provide another way to probe the end of the cosmological dark ages.

A difficulty with measurements of the REBL from fluctuations in the
near IR background is that they cannot be performed from ground-based
platforms; rapid variability in the near IR brightness a factor of
$\sim 10^{8}$ larger than the REBL signature arising from emission
from OH$^{-}$ in a layer reaching as high as $100 \,$km precludes
observations of resolved structures from the ground.  Teams working
with near IR capable space telescopes have begun detecting correlated
structure in the EBL consistent with a signature of REBL
(\citealt{Kashlinsky2005}, \citealt{Matsumoto2010}).  A feature of
these measurements is that they are at the limits of the ability of
the relevant instruments and require very complicated data analyses to
rectify the power spectra.  An instrument designed specifically with
the objective of measuring the fluctuation power of the diffuse EBL
would be able to mitigate some of these problems.

\section{The CIBER Instrument}
\label{S:ciber}

The CIBER instrument comprises four instruments which address three
separate but connected science cases outlined in Section
\ref{S:science}\ above; a description of each of these is presented
below.  More detailed descriptions of the instrument hardware and
implementation can be found in \citet{Bock2006} and \citet{Zemcov2010}.

\subsection{Imagers}
\label{sS:Imagers}

The CIBER Imagers are optimized to measure fluctuations in the near IR
EBL arising at the epoch of reionization.  Both Imagers are designed
and built to the same specification; the only difference between them
in flight is the choice of filter set.  A difficulty with previous
measurements of fluctuations in the near IR background is that,
typically, the field of view of the instrument has been small.  This
leads to the necessity of either measuring the fluctuations in very small
regions where shot noise from foreground galaxies is dominant or
mosiacking, which is technically difficult.  In either case, the
typical survey areas are too small to measure the angular scales at
which the fluctuation power from reionization is expected to peak.
The CIBER Imagers address this with a $2^{\circ} \times 2^{\circ}$
continuous field of view.  The use of commerically available near IR
HgCdTe detector packages with $1024 \times 1024$ pixels$^{2}$ lead to
a detector pixel size of $7"$ on the sky.  The Imagers also employ
cold shutters to measure the zero point of the detectors and
calibration lamps which allow production of a responsivity transfer
standard when in flight.

The Imager filter sets are interchangeable; for the first two flights
the bands were chosen to bracket the expected peak in wavelength of
the reionization fluctuation signal (examples of the expected color
of the fluctuations are shown in Figure \ref{fig:ebl}).  These bands
roughly corresponded to astronomical I and H bands, centered near $1.0$
and $1.6 \, \mu$m with $\lambda / \delta \lambda = 2$ for both
instruments.  Figure \ref{fig:irbps} shows the expected statistical
sensitivities of both of the CIBER Imagers compared to models of the
expected reionization and foreground power spectra in these bands, as
well as two bands longward at $2.2$ and $3.8 \, \mu$m.

\subsection{Low Resolution Spectrometer}
\label{sS:LRS}

The Low Resolution Spectrometer (LRS) is an $15 \leq R \leq 30$
spectro-photometric instrument designed to measure the near IR EBL at
$0.7 \leq \lambda \leq 2.1 \, \mu$m; a thorough description of the LRS
instrument can be found in \citet{Tsumura2010}.  The LRS comprises an
optical coupler with a 5-slit mask at the focus coupled to an imaging
camera with a $256 \times 256$ HgCdTe detector at its focus.  Spectral
dispersion is achieved using a prism located between the output of the
coupler and the camera.  The incident light is focused on the slit and
then recollimated by the coupler, dispersed by the prism, and imaged
onto the detector array.  The images of the 5 slits yield $5.^{\circ}
\times 2.8'$ cuts of the sky on one axis; the spectral dispersion
occurs on the other.  Masked parts of the array allow real time
monitoring of dark current and scattered light in the system.  As with
the Imagers, a cold shutter in front of the focal plane and a
calibration lamp allow monitoring of the detector offset and a
calibration transfer standard, respectively.

The LRS's sensitivity is $\sim 10 \,$\dnin\ pix$^{-1}$ in a 50 second
integration on the sky, meaning that in terms of raw sensitivity it
should be capable of reaching the surface brightnesses of the diffuse
signal from reionization when many pixels are averaged.  Of course,
even if the Zodiacal light were perfectly removed the LRS's imaging
capabilities are too modest to allow removal of foreground galaxies,
and the systematic errors associated with modelling the near IR EBL
would limit the overall sensitivity to the reionization component.
However, the LRS does show that modern IR detectors on small
components in suitably designed instruments could reach the
sensitivity limits required for absolute photometric measurement of
the REBL.

\subsection{Narrow Band Spectrometer}
\label{sS:NBS}

The final instrument in the CIBER payload is the narrow band
spectrometer (NBS).  This instrument is a fast refractive camera which
employs a relatively narrow band $R=1200$ filter; the geometry of the
instrument and filter causes the effective wavelength imaged at the
focal plane to change with position on the detector array.  The
absolute brightness of the ZL can be measured if the filter's central
bandpass is tuned to match a solar line which would be seen in
reflection off the Zodiacal dust, in the case of the first instrument
configuration the CaII line at 8542\AA.  Since slightly different
wavelengths are imaged to different parts of the focal plane, the
image yields the line depth directly.  As the absolute calibration
of the NBS is well understood and monitored (again using a cold
shutter and calibration lamp similar to the other CIBER instruments),
this line depth can be converted to physical units to determine the
absolute brightness of the ZL.  This is important as it removes the
need to model reflection of sunlight of the Zodiacal dust, which is a
difficult problem.  Furthermore, as the ZL is not expected to have
much structure over the NBS's $8^{\circ} \times 8^{\circ}$ field of
view, large $2' \times 2'$ pixels are sufficient to achieve the NBS's
science goals.  Such a measurement is not so important for the near IR
EBL fluctuation science, but is key for monitoring the ZL brightness
in the LRS data, and is also interesting data for the separate science
case of understanding the structure and composition of the Zodiacal
dust cloud near Earth.

\subsection{Instrument Package}
\label{sS:inst}

Together, the CIBER instrument package comprising the two Imagers, the LRS
and the NBS is designed to measure fluctuations in the REBL by comparing a
wavelength where they should be brightest to one shortward of the peak
and measure the absolute brightness of the EBL between $0.7$ and $2.1
\, \mu$m with real-time monitoring of the ZL emission in each science
field.  The four instruments are mounted to a common optical bench and
co-boresighted with the long axis of the rocket skin.  The optical
bench, instrument assemblies and detectors are cooled to $\sim 80 \,$K
using an on-board cryostat containing 7 litres of liquid nitrogen,
enough to keep the cryogenic insert cold for $\sim 18 \,$h.  Custom
cabling and warm electronics are used to clock and read out the
detector packages; the digitized detector readouts are then sent
onward to the NASA-built rocket electronics for transmission.  A
ground station is used to receive and condition the transmitted data
and store it to computer disk for later processing.  Great care is
taken in optically baffling the four instrument assemblies; the rocket
skin heats to $\sim 400 \,$K due to atmospheric drag on ascent and
stray thermal emission from the hot skin and rocket door can cause
drastically increased photocurrent at the detectors due to scattering
in the instrument.

The first CIBER configuration has deployed and flown two times from
White Sands Missile Range in New Mexico, USA using a Black Brant IX
sounding rocket vehicle; a description of the vehicle configuration
can be found in \citet{Zemcov2010}.  The first flight took place on
February 25, 2009 and yielded good results, though thermal
contamination from the skin caused much of the data analysis to be
extremely challenging.  Despite this, \citet{Tsumura2010} details this
discovery of a new, broad line in the reflected solar light from the
ZL which has implications for the source of the Zodiacal dust.
Particles from both asteroids and comets contribute to the Zodiacal
dust cloud; the overall albedo measured by CIBER shows that the
composition of the Zodiacal dust near the Earth is dominated by
minerals common in asteroids rather than in comets.

CIBER's second flight took place July 10, 2010 and yielded an
excellent data set.  Both flights observed a total of 6 science fields
which were divided up into 4 deep REBL fields, 2 ZL characterization
fields, and a separate calibration field during which the star Vega
was observed.  The REBL fields were chosen to have deep ancillary
coverage from instruments like IRAC and Akari to allow deep masking of
foreground galaxies down to $\sim 22 \,$mag in I-band.  Both flights
yielded $\sim 7$ minutes of astronomical integration time, and
achieved apogees $> 325 \,$km.  The REBL data analysis is currently
underway and as the data are very high quality it should be possible
to achieve the error estimates shown in Figure \ref{fig:irbps}\ with
them.

\section{The Future of Reionization Science from the Near IR EBL}
\label{S:future}

We conclude with a brief outlook for measurements of the absolute near
IR EBL and background fluctuations which will take place in the next decade.

\subsection*{The Future of CIBER}
\label{sS:ciber1}

The CIBER payload has been flown twice and is currently preparing for
a third flight, again on a Black Brant IX vehicle.  For its fourth and
final flight, CIBER will fly on a non-recovered Black Brant XII
vehicle which will achieve an apogee $> 1000 \,$km.  This vehicle is
an improvement both because of the increase in data collecting time in
space and because air glow from contaminants brought up with the
payload and near the Earth's atmosphere will be reduced.

Additional reconfigurations of the CIBER payload may include adding
polarization capability to the LRS to help disentangle the unpolarized
EBL from the polarized ZL, changing the solar Fraunhofer line to which
the NBS is tuned, or modifying the Imager bandpasses to optimize
measurement of the fluctuations from REBL.  These changes will be
informed by progress on the data analysis and instituted only if they
maximize the scientific output of CIBER's third and fourth flights.

\subsection*{CIBER-2}
\label{sS:ciber2}

In addition to the further flights of the CIBER payload, CIBER-2 is an
entirely new instrument package which will optimize sensitivity to REBL
fluctuations.  This will be achieved by increasing the aperture of the
telescope, thereby increasing the overall entendue of the system.
Additionally, the number of bands will be increased from 2 to 4, which
will allow a better measurement of the history of reionization.
CIBER-2 will leverage the technology heritage of CIBER to minimize
hardware development by reusing designs for components like the focal
plane assemblies, cryostat design, and warm electronics.  All told,
these changes should increase the sensitivity to fluctuations by a
factor of $\sim 10$ in all bands over that of CIBER.  This sensitivity
approaches the floor of the signal expected from reionization; if no
REBL signal were detected with CIBER-2, it would translate to very
strong constraints on minimally slow and weak reionization scenarios.

\subsection*{ZEBRA}
\label{sS:zebra}

Direct photometric measurements of the REBL will be challenging.
There are a number of bright foregrounds which need to be removed
before it will be possible to isolate the diffuse reionization
background itself: ZL, diffuse ISM emission, and emission from the
galaxies that comprise the EBL itself will need to be masked,
modelled, or otherwise mitigated.  Given the current state of the
measurements shown in Figure \ref{fig:ebl}, it might not ever be
possible to remove emission from the ZL to a level sufficient to
measure the REBL with direct photometry from near the Earth.  However,
the density of the Zodiacal dust falls sharply with helio-centric
distance; a naive extrapolation of the Pioneer Zodiacal dust counts
\citep{Hanner1976} suggests that, at $r_{\odot} > 5 \,$AU, the brightness of
the ZL should be a factor of $\sim 1000$ smaller than near Earth.
Clearly, deploying an instrument package to the outer solar system
would offer significant advantages for measurement of the (R)EBL
\citep{Cooray2009}. 

The Zodiacal dust, Extragalactic Background and Reionization
Apparatus\footnote{\tt{http://zebra.caltech.edu/}} (ZEBRA) has been
conceived as a concept for a high-readiness instrument which would
undertake science observations during the cruise phase of an outer
planet mission sometime in the next decade.  The ZEBRA concept calls
for two instruments working the optical and near IR, the Wide field
Absolute Module (WAM) and the High-resolution Absolute Module (HAM)
which would address a number of scientific objectives, including the
structure and composition of the Zodiacal dust cloud, a definitive
photometric measurement of the EBL, and a measurement of the
spectro-photometric properties of the REBL.  These observations would
be undertaken during the cruise phase of a host mission to the outer
planets; data would be available from $1$ to $\sim 10\,$AU, with the
bulk of the REBL data being taken at the largest heliocentric
distances.  If funded, ZEBRA would provide the definitive measurement
of the properties of the astrophysical sky between $0.5$ and $5 \,
\mu$m for a modest cost, and would allow us to address many questions
about the epoch of reionization.

The future of reionization science using near IR instruments is
bright; the next decade should bring interesting new constraints from
this field which will be complimentary to other methods of
investigation, and in fact allow unique constraints to be placed on
the history of star formation in the Universe.

\section*{Acknowledgments}
 
This work was supported by NASA APRA research grants (NNX07AI54G,
NNG05WC18G, NNX07AG43G, and NNX07AJ24G), and KAKENHI grants
(20$\cdot$34, 18204018, 19540250, 21111004, and 21340047) from the
Japan Society for the Promotion of Science and the Ministry of
Education, Culture, Sports, Science and Technology.  We acknowledge
the dedicated efforts of the sounding rocket staff at NASA Wallops
Flight Facility and White Sands Missile Range, and the engineers at
the Genesia Corporation for the technical support of the CIBER optics.
MZ acknowledges support from a NASA Postdoctoral Fellowship, KT
acknowledges support from the JSPS Research Fellowship for Young
Scientists, and AC acknowledges support from an NSF CAREER award.

\bibliography{ciber_desy}

\end{document}